\documentclass[conference]{IEEEtran}
\IEEEoverridecommandlockouts

\usepackage{cite}
\usepackage{amsmath,amssymb,amsfonts}
\usepackage{algorithmic}
\usepackage{graphicx}
\usepackage{textcomp}
\usepackage{xcolor}
\usepackage{hyperref}
\usepackage{makecell}
\def\BibTeX{{\rm B\kern-.05em{\sc i\kern-.025em b}\kern-.08em
    T\kern-.1667em\lower.7ex\hbox{E}\kern-.125emX}}
\begin{document}

\title{The path to 5G-Advanced and 6G Non-Terrestrial Network systems}

\author{\IEEEauthorblockN{Alessandro Guidotti\IEEEauthorrefmark{1}, Alessandro Vanelli-Coralli\IEEEauthorrefmark{2}, Vincenzo Schena\IEEEauthorrefmark{3}, Nicolas Chuberre\IEEEauthorrefmark{4},\\ Mohamed El Jaafari\IEEEauthorrefmark{4}, Jani Puttonen\IEEEauthorrefmark{5}, Stefano Cioni\IEEEauthorrefmark{6}}
\IEEEauthorblockA{\IEEEauthorrefmark{1}National Inter-University Consortium for Telecommunications (CNIT), Univ. of Bologna Research Unit, Italy}
\IEEEauthorblockA{\IEEEauthorrefmark{2}Dept. of Electrical, Electronic, and Information Engineering (DEI), Univ. of Bologna, Bologna, Italy}
\IEEEauthorblockA{\IEEEauthorrefmark{3}Thales Alenia Space Italia, Via Saccomuro, 24, 00131 Rome, Italy}
\IEEEauthorblockA{\IEEEauthorrefmark{4}Thales Alenia Space France, 26 Avenue JF Champollion, 31000 Toulouse, France}
\IEEEauthorblockA{\IEEEauthorrefmark{5}Magister Solutions, Sep\"{a}nkatu 14 C, 40720 Jyv\"{a}skyl\"{a}, Finland}
\IEEEauthorblockA{\IEEEauthorrefmark{6}European Space Agency - esa.int, ESTEC/TEC-EST, Noordwijk, The Netherlands.}
}

\maketitle

\begin{abstract}
Today, 5G networks are being worldwide rolled out, with significant benefits in our economy and society. However, 5G systems alone are not expected to be sufficient for the challenges that 2030 networks will experience, including, \emph{e.g.}, \emph{always-on} networks, 1 Tbps peak data rate, $<$10 cm positioning, etc. Thus, the definition of evolutions of the 5G systems and  their (r)evolutions are already being addressed by the scientific and industrial communities, targeting 5G-Advanced (5G-A) and 6G. In this framework, Non-Terrestrial Networks (NTN) have successfully been integrated in 3GPP Rel. 17 and it is expected that they will play an even more pivotal role for 5G-A (up to Rel. 20) and 6G systems (beyond Rel. 20). In this paper, we explore the path that will lead to 5G-A and 6G NTN communications, providing a clear perspective in terms of system architecture, services, technologies, and standardisation roadmap.
\end{abstract}


\section{Introduction}
\label{sec:introduction}
While 5G networks are already bringing benefits to all sectors of our economy and society, research and development efforts are already directed towards the design of enhanced 5G-Advanced (5G-A) features and the exploration of uncharted areas for future 6G communications, \cite{intro_01,intro_02,intro_03,intro_04}. 5G-A is expected to unleash the full potential of 5G by strengthening the network performance and by providing connectivity to all devices in all scenarios. Such enhancements will pave the way for the next generation of mobile communications, 6G. In 2021, ITU-R WP 5D initiated the development of the vision for IMT-2030 and beyond; moreover, the ITU-T Focus Group Technologies for Network 2030 (FG-NET-2030), between 2018 and 2020, defined a set of preliminary target services for 6G, \cite{itut_fg_net,itut_tech}. 6G systems are expected to create a fully connected world, with the convergence of the physical, human, and digital domains. The network will provide links between the domains through devices embedded everywhere, as well as the infrastructure and the intelligence of the digital domain. According to 5G-PPP, three classes of interactions will be possible, \cite{5gppp}: i) \emph{digital twinning} of systems, with sensors and actuators that can tightly synchronise the domains to achieve digital twins of cities, factories, or even our bodies; ii) \emph{connected intelligence}, with the network serving as the key infrastructure with trusted Artificial Intelligence (AI) functions managing virtual representations in the digital domain; and iii) \emph{immersive communications}, where high-resolution visual/spatial, tactile/haptic, and other sensory data can be exchanged with high throughput and low latency to create an immersive experience of being somewhere else. 

\begin{figure}[t!]
\centering
\includegraphics[width=0.4\textwidth]{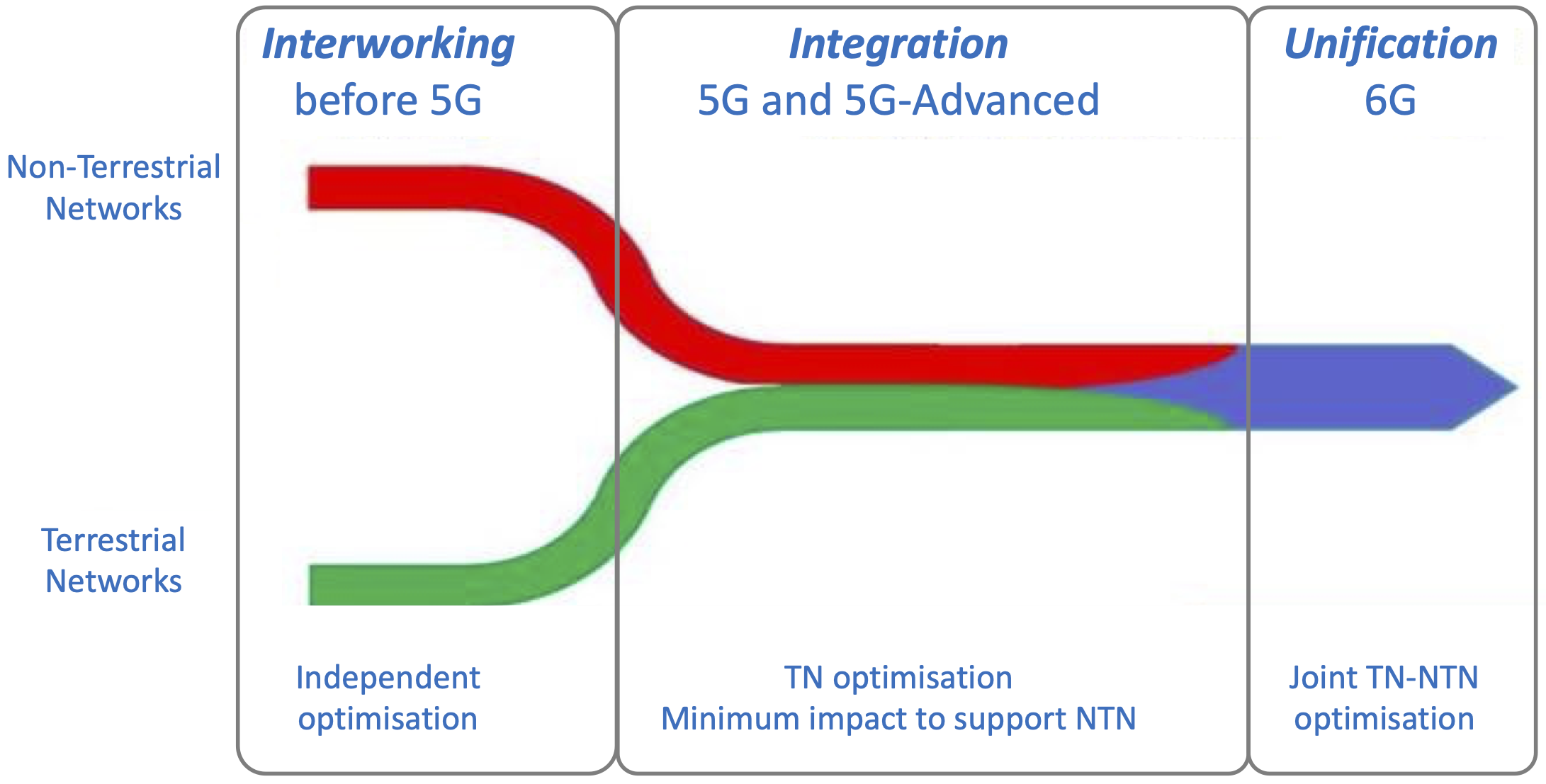}
\caption{Interaction between TN and NTN before and beyond 5G.}
\label{fig:path_to_6g}
\end{figure}

\begin{figure*}[t!]
\centering
\includegraphics[width=0.73\textwidth]{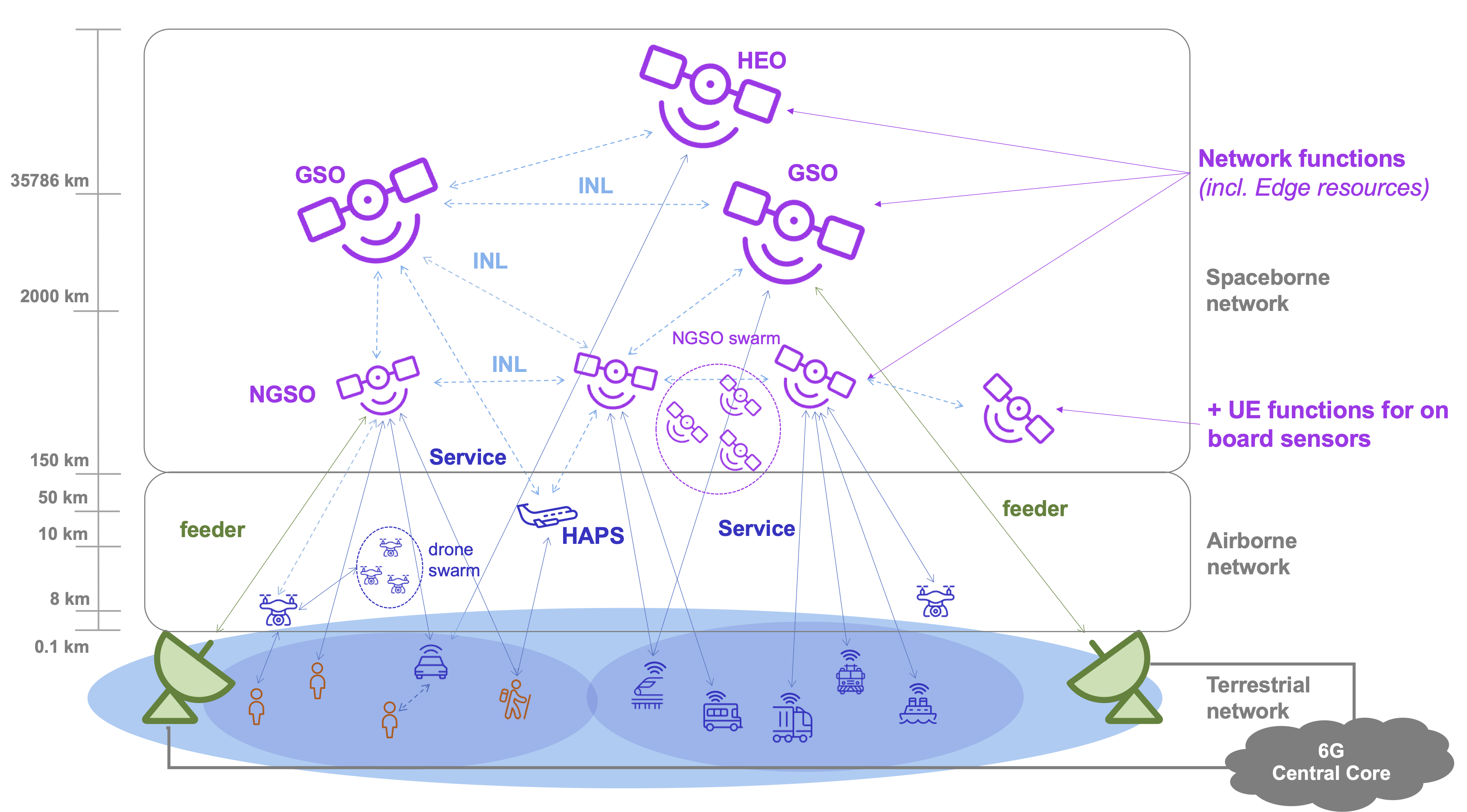}
\caption{The proposed multi-dimensional multi-layer architecture. [to be slightly adjusted]}
\label{fig:md_ml_arch}
\end{figure*}

In this framework, it is globally agreed that the full integration between terrestrial and non-terrestrial network components will be essential, \cite{intro_1,intro_2,intro_3,intro_4,intro_5,intro_6}. The added value of the integration of an NT segment in the New Radio (NR) architecture was recognised by 3GPP since Rel. 15. Two Study Items (SI) under Radio Access Network (RAN) and Service and system Aspects (SA) 1 related to the use of satellite access in 5G and the support of NR for Non-Terrestrial Networks (NTN) paved the way for the approval of the first dedicated NTN Work Item (WI), \cite{tvt_2019,cn_2021}. The outcomes of these SIs, in 3GPP TR 38.811, \cite{38_811}, provided a solid background in terms of use cases, scenarios, and characterisation of the NTN channel; in addition, this document also provided a preliminary analysis related to the impact of bringing the NR Air Interface and protocols on NTN links. In Rel. 16 two SIs were initiated in SA2 and RAN; building on the outcomes of Rel.15, TR 38.821, \cite{38_821}, reports the identified NTN architectures and a detailed analysis on the challenges and proposed solutions related to Layer 1 (\emph{e.g.}, physical layer procedures, including Random Access and Timing Advance), radio protocols (\emph{e.g.}, HARQ, mobility management in the Control Plane), and architecture and interfaces (\emph{e.g.}, tracking area management, management of the network identities). The above studies represented a turning-point for the definition of a truly integrated NT component in the terrestrial 5G system, starting from the recently completed Rel. 17, which also addressed a SI on Internet of Things (IoT) via NTN. As of today, the analyses related to NTN in 5G systems have been mainly oriented towards a thorough feasibility assessment, aiming at identifying the required adaptations for NR techniques and technologies allowing the exploitation of NTN links based on State-of-the-Art (SoA) space technologies and industrial assets. In Rel. 17, the solutions to tackle the issues related to long propagation delays, large Doppler shifts, and moving cells in NTN were specified. The main objective was that of adapting the existing specifications, which were designed for terrestrial networks (TN), to NTN. Many enhancement were introduced to support 5G NR NTN and IoT NTN based radio access. The main focus was on transparent payload based on Low Earth Orbit (LEO) and Geostationary Earth Orbit (GEO) network scenarios addressing 3GPP power class 3 User Equipments (UEs) with Global Navigation Satellite System (GNSS) capabilities in both Earth fixed and moving cell configurations; in terms of duplexing, Frequency Division Duplexing (FDD) was addressed. The main targeted NTN deployment scenarios are: i) 3GPP-defined NTN access network providing direct connectivity in LEO/MEO systems operating in FR1 (4.1-7.125 GHz) and addressing handset devices with mobile broadband services; ii) 3GPP-defined NTN access network providing direct connectivity based on GEO and LEO/MEO systems operating in sub-FR1 and addressing IoT devices with narrowband service (NB-IoT/eMTC based radio access); and iii) 3GPP-defined NTN access network providing indirect connectivity, which will be supported in Rel. 18, in GEO and LEO/MEO systems operating in Ku or Ka band (\emph{i.e.}, above 10 GHz) and addressing Very Small Aperture Terminal (VSAT) type devices (fixed and vehicle mounted) with mobile broadband services, \cite{rel_18_items}. The NTN journey in the 5G ecosystem will continue in Rel. 19 and beyond. Many other features are being discussed as potential enhancements to be introduced as part of 5G-Advanced, including, but not limited to, the support of regenerative payloads and further enhancements for capacity and coverage improvement, \emph{e.g.}, Multi-Connectivity (MC), Carrier Aggregation (CA), optimisation of the downlink Peak-to-Average Power Ratio (PAPR).

In this context, while the interest and effort in NTN has been continuously increasing, it is clear that further enhancements will enable better performance and/or new capabilities for the NTN component Beyond 5G (B5G), \emph{i.e.}, in 5G-A and 6G systems, \cite{intro_7,intro_8,intro_9,intro_10}.  As shown in Fig.~\ref{fig:path_to_6g}, before 5G, TN and NTN were independently optimised; then, with 5G and 5G-Advanced, the objective has been the optimisation of the TN and integration of the NTN component with minimum impact. However, only with 6G systems TN and NTN will be jointly optimised in a unified and fully integrated multi-layered infrastructure. Such architecture will combine terrestrial, airborne, and spaceborne radio access networks for the envisaged convergence of the physical, human, and digital worlds. 

In this paper, we provide a detailed overview on the technologies and services targeting highly efficient and deeply integrated satellite networks in 5G-A and 6G cellular systems; to this aim, a the multi-layer multi-dimensional NTN architecture is proposed. The main research and innovation trends are discussed also highlighting the standardisation roadmap towards Rel. 20 and beyond. 

\section{The NTN multi-band multi-layer multi-dimensional architecture}
\label{sec:architecture}
In order to satisfy the performance requirements of 5G-A and 6G systems and to foster the connection among the physical, digital, and human domains, a multi-band multi-layer multi-dimensional (MB-ML-MD) network architecture will be a fundamental part of future infrastructures, \cite{intro_6}. As shown in Fig.~\ref{fig:md_ml_arch}, the bi-dimensional terrestrial infrastructure is augmented by a third dimension (multi-dimensional) provided by an NT component, consisting of airborne and spaceborne communication nodes at different orbits (multi-layer), operating in FR1 (L, S, C bands) and FR2 (Ku and Ka bands). The nodes in the NT segment of the MB-ML-MD architecture can communicate with each other by means of Inter-Node Links (INLs), with a nomenclature further remarking that NTN is composed not only by satellites but also by High Altitude Platform Systems (HAPS), drones, and Unmanned Aerial Vehicles (UAVs). Such INLs can be intra-orbit and inter-orbit. Notably, depending on the specific platform type of the flying node, different characteristics and benefits are provided to the global network: i) Geosynchronous Orbit (GSO) satellites provide fixed continental coverage with large propagation delays, which can efficiently support non delay sensitive and broadcasting communications; ii) Non GSO (NGSO) satellites provide a moving service area (potentially global, with a sufficiently large constellation), with low latency at low orbits (down to a few ms), which can support delay sensitive applications; iii) HAPS can provide a quasi-static regional coverage with a latency comparable to that of terrestrial large cells, making them fit for regional hot spot scenarios; and iv) UAV and drones can provide coverage to specific small areas tailored to the users' needs, which is attractive for highly reliable and low latency broadband access. Based on the proposed architecture and the above observations, below we thoroughly review the most relevant services and technologies for 5G-A and 6G communications.

\section{NTN services}
\label{sec:services}
The unification of satellite and terrestrial networks allows to conceive reliable, secure, and cost-effective communication services. In 3GPP TR 28.822, \cite{22_822}, a list of potential services for 5G systems incorporating an NT component is provided mainly covering the enhanced Mobile Broadband (eMBB) and massive Machine Type Communication (mMTC) verticals. Below, we briefly summarise and elaborate the most interesting that can be considered to be relevant also for 5G-A and 6G. 
\paragraph{Broadcast/multicast through satellite} With an ever increasing capacity request (\emph{e.g.}, due to the increase in the number of Ultra-High Definition (UHD) programs in broadcasting services), NTN can provide efficient access to: i) serve users located in un-served areas; and ii) serve users with the required Quality of Service (QoS) when the Mobile Network Operator (MNO) is saturating due to the large traffic requests, \emph{i.e.}, for traffic off-loading.
\paragraph{5G to premises} A terrestrial operator aims at gathering the generated traffic at an office gateway located in the premises area. Broadcasting and multicasting services will be provided over the NTN component; in addition, if latency is not critical, it can also be used for unicast communications, thus off-loading the traffic from the terrestrial infrastructure.
\paragraph{Emergency management} This service refers to situations in which a natural disaster or a terrorist attack destroy, fully or partially, part of the RAN or 5G core. Consequently, all of the services provided by one or more MNOs operating through the disrupted terrestrial infrastructure cannot be guaranteed anymore. The restoration of the communication infrastructure on the area is fundamental for both the population and, in particular, the first responders to coordinate their efforts and to report to the command center the evolution of the rescue operations. In this scenario, both LEO satellites or HAPS can provide coverage with a sufficiently low latency. It is worth mentioning that the capacity requests in such scenario can range from very low values (\emph{e.g.}, simply reporting the locations of the different rescue teams to the command center, in the same area or remote) to quite large (\emph{e.g.}, Augmented Reality helmets provided to the first responders, sending their videos to the command center).
\paragraph{IoT via satellite} Many potential services can be provided for IoT through NTN, \emph{e.g.}: i) global NB-IoT/eMTC coverage, guaranteeing a continuous global coverage of NB-IoT/eMTC devices for any type of data transfer between the terminals and a central server, as long as non delay critical communications are involved; ii) remote control/monitoring of critical infrastructures, \emph{e.g.}, off-shore wind farm; and iii) smart good tracking, to continuously and seamlessly monitor and localise goods transported by ship, train, or cargo flight. Looking at 5G-A and 6G, to best exploit the benefits of NT infrastructures and cope with their limitations, the following NT elements should be considered: UAVs and drones, for on-demand coverage of small areas for limited amounts of time, such as in the event of an emergency causing ground based infrastructure to malfunction or become unavailable, and HAPS, stationary or quasi-stationary airships at an altitude of 20 km offering regional coverage. In this framework, the following services can be envisaged:
\paragraph{Drones and UAV for disaster relief} As mentioned, terrestrial infrastructures are vulnerable to natural or man-made disasters and their damaging increases the difficulties that first responders may encounter. Today, communications via satellite may be provided by using civilian satellites and satellite-specific ground terminals, both handheld and fixed, but they may not always be available. Additionally, those terminals are often bulky and unavailable to non-relief personnel. An evolution could be the use of deployable support drones and UAVs to cover a large area and offer backhauling services for end users: contrary to current systems, handheld terminals should be able to interface and directly access to the drones/UAVs, allowing a larger number of users to access the network. The drones/UAVs act as Access Points (APs), providing a relay for the on-ground users towards LEO satellites. From there, an Inter-Satellite Link (ISL) network could route the data to the various core network gateways (GWs). Such network would be temporary and require Software Defined Network (SDN) and Network Function Virtualisation (NFV) technologies for best scalability and agility. SDN and NFV will play an important role in the shift to network slicing. Virtualisation will enable separation of the software from the hardware and offer the possibility to instantiate many functions on a common infrastructure. With this approach, the infrastructure can be shared by different tenants and provide different services.
\paragraph{GEO/LEO/HAPS for trains/planes/ships} To provide Internet connectivity to the passengers, existing infrastructures based on Wi-Fi are often underperforming and unreliable, while the ground network are often in underserved areas. The small size of the ground 5G cells, comparable at most to 4G coverage, would also cause numerous handover requests as the vehicle passes through various cells. A MB-ML-MD NTN coverage would be a fitting solution by providing a global service continuity and resiliency: a combination of TN and NTN can provide service continuity for such use scenario and higher reliability/availability. In this case, high orbits can be used to guarantee a wide and global coverage, while low orbits provide larger capacity and lower latency. A regenerative MB-ML-MD NTN can host the 5GC functions, including the User Plane Function (UPF). The UPFs can be deployed on-board and with a distributed approach, close to the users to reduce the latency for delay-sensitive applications. On-board cloud capabilities can support both the radio access and core functions for scaling and bringing services to the mobile edge. HAPS offer the opportunity for wide area access: their altitude allows for direct access by handheld devices, as well as the use of large antennas, and stationary coverage, collecting data on a large area reducing the impact of handover for moving vehicles. Additionally, their stationary flight allows for a noticeable reduction of Doppler issues, as only the vehicle would be moving. Their size allows to mount directive active antennas able to follow a LEO satellite in flight, granting a longer visibility, which can be further improved through ISLs. Low latency use cases can also be considered, taking into account the possibilities of ISLs; however, including the rerouting and processing delays, such an architecture cannot be used for remote control or navigation. Also in this case, an autonomous rerouting capability would be necessary: SDR and shared resources to optimise the connectivity between HAPS and LEO/GEO depending on the typology of services would be fundamental, allowing the network to act as a terrestrial network, with access nodes and smart rerouting of information depending on the requested latency and QoS. Finally, in the context of moving terminals, it is worthwhile mentioning that the 5G Automotive Association (5GAA) recently opened a WI in 3GPP for V2X services via NTN.

\paragraph{HAPS for non-connected areas} Similar in scope to the previous proposal, various areas with low population density, such as mountain communities, small islands, agricultural regions, as well as mining and lumbering facilities, require connectivity. Their remoteness, as well as legal or environmental limitations, or geographical issues may contribute to the unfeasibility of a ground-based infrastructure. As of today, coverage can be obtained through the use of satellite-based connectivity, requiring specialised terminals, antennas, and having to deal with the limitations of direct satellite access: increased Doppler and limited visibility time for LEO, and increased latency and noticeable attenuation for GEO.
A MB-ML-MD NTN composed of HAPS covering the areas of interest, and linking to LEO/GEO nodes via ISL, would offer a solution to such shortcomings: HAPS have the advantage of acting as ideal APs to the network. Their altitude allows for direct access from ground users, and their stationarity and wide area coverage allows to reduce handover and pointing issues. Connectivity to GEO or LEO satellites would allow for large throughput, high latency services, or small throughput, low latency services as well. A mesh topology is indeed the most compelling option, with three layers: HAPS granting access to the network, LEO satellites acting as network nodes and rerouting low latency services, and GEO satellites acting as high throughput backbones. This option can also be implemented in already served areas in case it would be necessary to offload some of the traffic from the ground infrastructure. In all those cases, similarly to the previous scenarios, it is fundamental to provide SDR capabilities for rerouting and smart resource allocation, so as to optimise the network resources.

\section{Technologies and standardisation roadmap}
\label{sec:technologies}
Leveraging Rel. 17 NTN specifications, and targeting the above services, both evolutions of the available technologies (5G-A) and revolutionary concepts (6G) shall be developed.
\subsection{5G-Advanced technologies}
In Rel. 18, for mMTC services the enhancements will be directed towards improving the performance in discontinuous coverage and deal with terminal mobility. As for eMBB, the following features will be addressed: i) network-based UE location determination; ii) coverage enhancements; iii) NR-NTN deployment above $10$ GHz and support for VSAT/ESIM terminals; iv) NTN-TN and NTN-NTN mobility and service continuity enhancements. Then, for 5G-A, some of the techniques de-prioritised in Rel. 18 can be evaluated as candidate features for Rel. 19 and beyond.

\paragraph{NTN-NTN asynchronous MC and CA} MC and CA are viable approaches to increase the user throughput and Quality of Experience (QoE) by means of spatial diversity between NGSO-NGSO and GEO-NGSO satellites, \emph{i.e.}, satellites at potentially very different orbits as in the introduced global architecture. CA, introduced in LTE, is a technique in which multiple carriers are aggregated to serve the same UE; on the other hand, MC allows the same UE to be connected to multiple nodes simultaneously, increasing the transmission robustness and reliability. Intra-satellite CA is particularly important in high frequency bands to provide very large throughput or to increase the flexibility with frequency channel planning: smaller channels can improve the link budget and they can be aggregated if larger bandwidths are required. In this framework, the different propagation delays and/or network topologies between the various access nodes pose a challenge and shall be taken into account. Moreover, the system shall also handle different time and frequency compensations, as well as the optimised selection of the master node versus the secondary node. 

\paragraph{Beam management and Bandwidth Part (BWP) association enhancements} the procedures for data-driven beam management and BWP optimisation should be revisited. Since BWP is the primary method to partition the carrier bandwidth and accommodate multiple numerologies and UE types with different downlink capabilities, such flexibility should be preserved for the operator, on top of allowing the use of BWP to manage multiple beam configurations and arbitrary frequency reuse schemes. Beam determination techniques would eventually rely on historical traffic data and satellite ephemeris. It would be relevant to understand if the legacy beam procedures can support this for certain NTN scenarios. BWP flexibility for channel partitioning should be preserved and possibly extended (\emph{e.g.}, a larger number of maximum BWPs) on top of its flexible use for NTN beam and frequency reuse management. Since it is not yet clear whether Rel. 17 beam management procedures can already support this, these aspects shall be addressed in the path to 5G-A.

\paragraph{Complete support for HEO and MEO and hybrid multi-orbit architectures} NTN should cover all satellite orbit designs, to avoid limiting the operator choices. In particular, Highly Elliptical Orbit (HEO) was omitted from Rel.17 and it should be addressed since real systems are being deployed there. Medium Earth Orbit (MEO) was included as a parameter set and likely implicitly covered by GSO and LEO specifications, but it should be explicitly consolidated. NTN systems are expected to include multiple satellites in different orbits, as shown in the MD-ML architecture in Section~\ref{sec:architecture}. In this framework, the following aspects will be addressed:
\begin{itemize}
	\item include HEO/MEO orbit parameters in NTN specifications and guarantee that they can actually support them;
	\item HEO Doppler is more benign compared to LEO, but a Doppler variation is expected due to the orbit eccentricity. The maximum path delay is expected to be slightly higher than GEO (it is to be clarified if the current worst total path delay is sufficient or not) due to HEO orbit apogee being higher than GEO;
	\item MEO is already implicitly covered by the conjunction of GSO and LEO, so  it is not expected to require a significant effort. However, some example parameters should be captured in the specifications for completeness;
	\item MEO was just implied in TR 38.821, \cite{38_821}. Thus, it needs to be explicitly captured and specified (leveraging the conjunction of GSO and LEO).
\end{itemize}

\paragraph{Support of Multicast and Broadcast Service (MBS)} Supporting MBS would introduce benefits in several services, \emph{e.g.}, public warning service, distribution of software upgrades, multimedia content delivery, and IoT applications requiring the triggering or distribution of messages. The following aspects shall be taken into account: i) multicast transmissions accommodating extra delays, including HARQ de-activations; ii) addressing broadcasting service continuity issues in NGSO when targeting Earth fixed coverages; and iii) multicast/broadcast should take into account the beam topology and both Mobile-Originated and Network-Originated multicast.

\paragraph{UE without GNSS} The energy efficiency of the UEs can be improved by reducing the dependency on the GNSS service availability, providing support for low cost UEs. Notably, this calls for enhancements related to new methods for uplink time and frequency synchronisation in idle and connected mode, new PHY Random Access Channel (PRACH) specifications, and assessing the impact related to the support of UEs with and without GNSS in a given cell.

\paragraph{High performance UE} The link budget of the UE, in particular on the uplink, can be improved for specific terminal classes, \emph{e.g.}, first responders or drones, by means of circularly polarised antennas so as to reduce the polarisation losses. Moreover, such advanced receivers might also be equipped with better noise figure elements, so as to improve the forward link budget. Higher antenna gains or larger transmission power levels can also be considered.

\paragraph{Relay-based architecture for NTN} A VSAT/ESIM should be able to act as an Integrated Access and Backhaul (IAB) node, with the NTN-gNB acting as IAB-Donor. This architecture is useful to support multiple V2X and hot-spot configurations. In this context, the extension of IAB support to NTN (VSAT/ESIM acting as relay) might be the best scope to start, but other options are possible. The IAB part is what is influenced by the NTN link. Other local relay functions (\emph{e.g.}, V2X, sidelink, etc.) have less relevance specifically to NTN.

Other technologies that can be of interest for 5G-A are: i) Protocol simplification for Voice over NR (VoNR), aiming at increasing the user goodput, in particular in direct access to handheld terminals, the signalling overhead can be reduced; ii) support reduced NR bandwidth, aiming at improving the Maximum Coupling Loss (MCL), in particular on the downlink, channel bandwidths below 5 MHz can be considered, which require NR signalling adaptations; iii) NTN support for IoT, aiming at allowing a single core and radio architecture to support all services, including NTN IoT solutions; and iv) Half Duplex (HD) FDD, aimed at simplifying  the UE architecture (no need for duplexer) and the support of a larger spectrum to cover all NTN bands below 6 GHz, also leveraging the principles of Reduced Capability (RedCap) terminals.

\subsection{6G technologies}
The following technologies are considered as the building blocks that will allow to achieve the full and seamless integration of terrestrial and NT infrastructures in 6G. It shall be noticed that some of these technologies can already be preliminary introduced within 5G-A, \emph{i.e.}, there is not an abrupt separation between 5G-A and 6G, with the former rather being a bridge between legacy 5G and future 6G systems. 
\paragraph{Regenerative payloads and active antennas} Satellite payloads are now equipped with flexible On-Board Processors (OBPs), which provide advanced capabilities to improve the link budget and response times, as well as allowing to move core network features on-board by means of SDN capabilities and functional split. In this context, the payload is actually part of the global network, \emph{i.e.}, it allows the possibility to implement processing in the lower layers (gNB-DU), higher layers (full gNB), network functions as core network (\emph{e.g.}, User Plane Function) or even edge computing, and in-space routing by also exploiting ISLs. For the latter, it shall be noticed that ISLs can be between two satellites of the same constellation (intra-/inter- orbital plane), between two satellites of different constellations/orbits (\emph{e.g.}, between LEO and MEO satellites), or even between a satellite and a terrestrial node. Thus, advanced routing schemes taking into account the global 3D network topology are required. This calls for the design and deployment of more advanced OBPs and one of the challenges to be faced is related to the potentially huge dimension of routing tables.

The advanced processors, combined with active antenna arrays, offer the possibility to continuously optimise the resource management by allocating the power/time/frequency/space resources to where they are required by means of user-centric (Cell-Free) beamforming. The implementation of these techniques in NTN is non-trivial due to the need for accurate Channel State Information (CSI) at the transmitter. This challenge is even more critical assuming mobile UEs, \emph{e.g.}, on fast moving platforms as airplanes or trains, and NGSO systems, with fast moving satellites. To tackle these challenges, the following solutions can be envisaged: i) the availability of advanced OBPs would allow to compute the beamforming coefficients on-board, thus significantly reducing the impact of latency and CSI aging, \cite{{lb_precoding}}; ii) assuming that the UEs have GNSS capabilities, the CSI might be predicted by combining such information with the satellite ephemeris and/or by exploiting Machine Learning (ML) algorithms. The implementation of location-based, rather than CSI-based, solutions would also allow to avoid potentially complex channel estimation procedures at the UEs, in which it might be difficult to estimate some of the coefficients due to the large C/I at the receiver. When considering NGSO systems, in particular LEO mega-constellations, it shall be noticed that multiple satellites can be seen from the UE location (even tens of satellites). In this context, distributed beamforming can be implemented in order to also exploit the spatial diversity provided by dropping the usual assumption of co-located radiating elements. Such solutions can also be envisaged for HAPS connectivity. Another aspect to be taken into account is related to the actual traffic demand from the users, which can be significantly non-uniform across the service area. In such scenarios, regular beam lattices obtained with a pre-defined value of the radiation pattern at the beam edge might lead to a non-optimal resource distribution; by fully exploiting regenerative payloads and active antenna arrays, the OBP might be able to generate narrower beams in high traffic areas and wider beams in areas where the overall traffic request is more limited, \emph{i.e.}, to tailor the resource allocations to the on-ground ``hot'' and ``cold'' spots. In this framework, also mixed frequency reuse schemes can be envisaged, \emph{i.e.}, cold spot areas to be served with 3 or 4 colour schemes and hot spot areas in full frequency reuse. 

\paragraph{AI and ML} Due to the potentially large speed of the satellites (in particular LEO/VLEO) and to the heterogeneity in the global network due to the deployment of satellites on multiple orbits, aerial elements (drones/HAPS/UAV), and ground elements, the system optimisation shall be performed in a highly dynamic environment. AI and ML are widely recognised as the most promising solution in such dynamic and information-rich contexts in wireless communications. An interesting review on possible applications of AI/ML concepts to NTN is provided in \cite{ai_ml} and the references therein. Among these, Radio Resource Management (RRM) algorithms, including beamforming and Beam Hopping (BH), are one of the most likely applications. With respect to beamforming, AI solutions can be related to two main aspects: i) channel estimation, as discussed above; and ii) scheduling. Notably, scheduling in beamformed NTN systems is non-trivial, since the UE performance can only be known after the beamforming matrix has been computed, \emph{i.e.}, after the scheduler selected the UEs to be served; thus, this information cannot be exploited \emph{a priori} to optimise the scheduler and often iterative (computationally intensive) solutions need to be used. In this framework, supervised (regression) or unsupervised (clustering) ML algorithms or Deep Neural Networks (DNN) can be evaluated to identify the best scheduling options based on ancillary information, \emph{e.g.}, user location and/or traffic requests and satellite ephemeris. BH emerged as a promising technique to achieve a significant flexibility in adjusting the capacity provisioning to the traffic requests, in particular in traffic-driven implementations. The optimisation of the BH illumination plan is often formulated as an optimisation problem, to be solved by means of Genetic Algorithms (GA) or heuristic iterative solutions. The major challenge in such approaches is related to the identification of the global optimum instead of local optima when the search space is large, \emph{i.e.}, with a large number of beams. Recently, aimed at overcoming such limitations, AI algorithms have been proposed showing that multi-objective optimisations can be obtained for resource allocations. As already mentioned, accurate channel estimates are fundamental in RRM algorithms and to assess the system performance in order to optimise its design. In this framework, some recent studies have exploited satellite/aerial images to derive approximate channel parameters, such as the standard deviation of the shadowing loss and the path loss exponent or even the Reference Signal Received Power (RSRP). When image-based algorithms are implemented, a typical approach is that of exploiting Convolutional NN (CNN). The implementation of image-based solutions might be not feasible for NGSO systems, due to the rapidly changing environment for each satellite which would require a huge amount of images to be analysed in real-time, thus also introducing a significant overhead. In this framework, the possibility to implement linear/non-linear regression algorithms and NNs can be considered to predict/approximate the channel coefficients based on ancillary information, such as the user location, the satellite ephemeris, antenna models and configuration, gyroscope data, etc. AI solutions can be also implemented aimed at detecting the ionospheric scintillations, forecasting of network traffic, remote sensing, and telemetry mining, among the others.

\begin{table*}[t!]
		\renewcommand{\arraystretch}{1.2}
		\caption{NTN Study and Work Items in 3GPP Rel. 18.}
		\label{tab:3gpp_rel_18}
		\centering
		\begin{tabular}{|c|c|c|c|}
			\hline
			\textbf{Item} & \textbf{Lead} & \textbf{Title} & \textbf{Completion}\\
			\hline\hline
			FS\_5GET ``Extra territoriality'' & SA1 & Guidelines for extra-territorial 5G Systems (5GS) & Dec. 2021\\
			\hline
			WI ``NR\_NTN-enh'' & RAN2 & \makecell{Enhancements to Solutions for NR to \\ support non-terrestrial networks (NTN)} & Dec. 2023\\
			\hline
			WI ``IOT\_NTN-enh'' & RAN2 & \makecell{Enhancements to Solutions for NB-IoT \\ \& eMTC to support non-terrestrial networks (NTN)} & Dec. 2023 \\
			\hline
			SI ``FS\_5GSAT\_ARCH\_Ph2'' & SA2 & 5GC enhancement for satellite access Phase 2 & Jun. 2023 \\
			\hline
			SI ``FS\_5GSATB'' & SA2 & Study on satellite backhauling & Jun. 2023 \\
			\hline
			SI ``FS\_eLCS\_ph3'' & SA2 & Enhanced location services & Jun. 2023 \\
			\hline
		\end{tabular}
\end{table*}

\begin{figure*}[t!]
\centering
\includegraphics[width=0.65\textwidth]{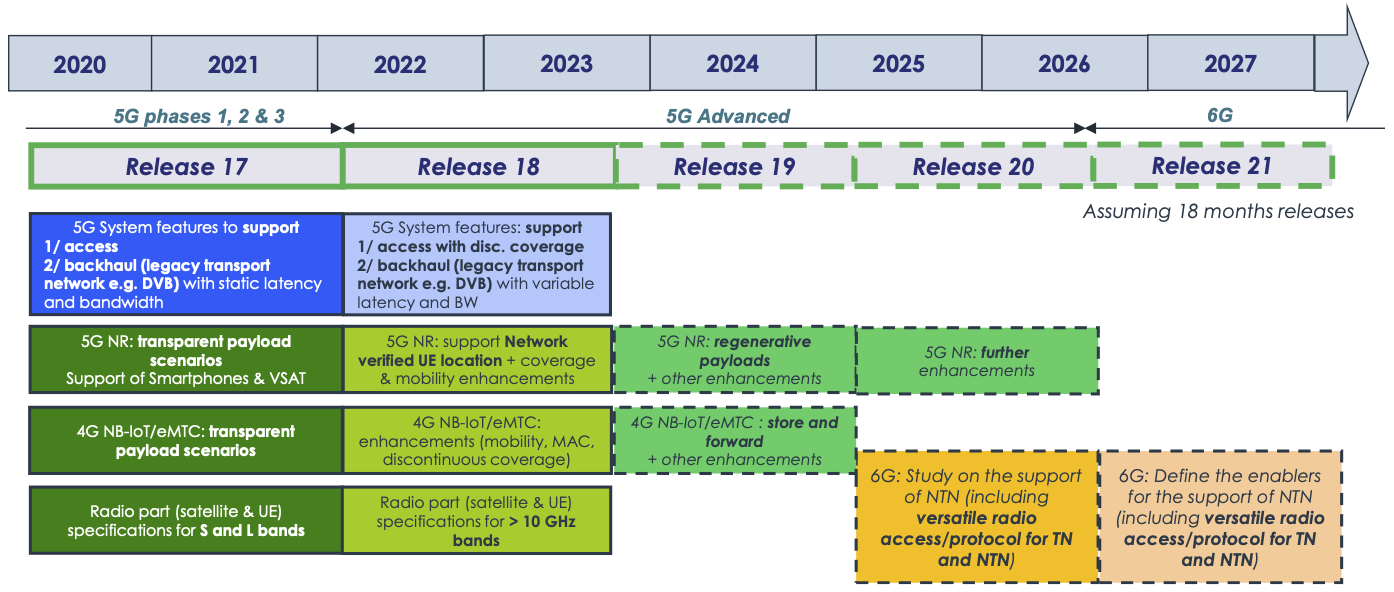}
\caption{NTN 3GPP roadmap.}
\label{fig:3gpp_timeline}
\end{figure*}

\paragraph{Next generation waveforms} High PAPR causes significant power efficiency and, thus, throughput losses; this aspect was already reported in Rel. 17, but for the moment being, despite its critical importance in terms of commercial viability and system capacity, it has not yet been addressed. This issue has been identified both above and below 6 GHz, but due to typical payload architectures it is more significant above 10 GHz. In particular, in Rel. 17 it was identified that the efficiency of power amplifiers decreases for increasing frequencies towards 100 GHz, \cite{wave_1}. Moreover, for transponder configurations with a small number (\emph{i.e.}, less than 3-4) of OFDM channels per High Power Amplifier (HPA), the throughput degradation may be severe. In this context, it is also worthwhile highlighting that PAPR alone is not a sufficient metric. The Out-Of-Band-Emission (OOBE) is also a significant player even in configurations with a large number of carriers per HPA, where the carrier PAPR is less problematic. In \cite{wave_2}, numerical simulations were performed to evaluate the performance degradation with a single carrier per transponder. Such losses are in the 2.5-6.5 dB range; the analysis in \cite{wave_2} also shows that the total degradation is independent of the selected modulation and amplifier type. It should also be noticed that the high power fluctuation of the CP-OFDM waveform affects all payloads in the forward link due to gain compensation/compression that may be necessary to normalise the PAPR to some extent even at lower frequency bands. In this framework, the uplink already enables DFT-s-OFDM which features very low PAPR requirements and hence is comparable to the DVB-RCS2 Multi-Frequency Time Division Multiple Access (MF-TDMA) waveform; however, it is not sure whether this will be implemented in mmWave bands. Some preliminary analyses showing the performance of different variants of the OFDM waveform, namely DFT-s-W-OFDM, W-OFDM, DT-s-f-OFDM, and f-OFDM, are reported in \cite{wave_3}, with different numerologies and OBO values. Here, it is shown how the different numerologies significantly impact the performance; moreover, it is also shown how large subcarrier spacings can actually accommodate the large Doppler shifts in NGSO systems. In addition to PAPR and OOBE, the following enhancements can be considered at waveform level: i) reduced PRACH format to allow multiple PRACH transmissions in the same beam as well as extended link margin; ii) non-orthogonal schemes to increase the number of terminals that can be simultaneously served; and iii) Coordinated Multi-Point (CoMP) transmissions to increase peak user data rate.

\paragraph{Reflecting Intelligent Surfaces (RIS)} A RIS is a thin meta-surface integrated with passive electronic components/switches that allows to control and manipulate the wireless signals that arrive on its surface, \cite{ris_1}. They can be used to modify the amplitude/phase of the signals, thus allowing to re-direct it towards desired directions. In terrestrial networks, the exploitation of RISs has been extensively assessed, while its application to NTN is still in its infancy. In \cite{ris_2}, link budget aspects for aerial platforms implementing RIS are discussed, with both far-field and near-field models. In \cite{ris_3}, the focus is on refracting RIS, in which all incident signals can pass through the surface, allowing to reconstruct the signals sent from the satellite indoor, which would be otherwise blocked. Finally, in \cite{ris_4}, the implementation of RIS-assisted communications in GEO systems is proposed; in particular, this system envisages both a direct link and a RIS-reflected link arriving at the UE; a joint optimisation problem is formulated to define the optimal power allocation and reflecting phase shift design.

\subsection{Standardisation roadmap}
The satellite component included in Rel. 17 made the integration of an NT component with the mobile systems possible. This standard is the result of a joint effort between stakeholders of both mobile and satellite industry who both find benefits: i) satellite operators can access a unified and large eco system and drive down the cost through economy of scale; and ii) mobile systems can achieve global service continuity and resiliency. In Q2 2022, several 3GPP SIs and WIs started targeting Rel. 18. These studies, summarised in Table~\ref{tab:3gpp_rel_18}, provide the first step into 5G-A. Fig.~\ref{fig:3gpp_timeline} shows our vision of the roadmap for NTN standardisation covering up to Rel. 21. It shall be noticed that the items related to Rel. 19 and above are not yet defined in 3GPP and they represent an educated guess by the authors. The roadmap is in line with the technologies previously discussed for 5G-A and 6G.

\section{Conclusions}
\label{sec:conclusions}
5G networks are being deployed all over the globe and the economy and society are already benefitting from the provided services. To fully unleash their potential, 5G-A systems are already being designed; moreover, targeting further enhanced capabilities and the convergence of the physical, human, and digital worlds, the definition of the 6G vision has begun. In this framework, leveraging the success of their integration in Rel. 17, it is globally recognised that NTN will play a pivotal role in future fully integrated systems. In this paper, we proposed a vision for a future MB-MD-ML integrated network, in which terrestrial, airborne, and spaceborne nodes strictly cooperate. The services that can be provided through 5G-A and 6G systems have been discussed, together with the enabling technologies that will be the building blocks of such infrastructure. For the identified technologies, a possible standardisation roadmap in the context of 3GPP is also discussed.

\section*{acknowledgements}
This work has been carried out in the framework of the EAGER (tEchnologies And techniques for satcom beyond 5G nEtwoRks) project, under a programme funded by, the European Space Agency (ESA). The views expressed in this document can in no way be taken to reflect the official opinion of the European Space Agency.

\vspace{12pt}

\end{document}